\documentclass[aps,prd,amsmath,twocolumn,amssymbaps,showpacs]{revtex4-1}
\usepackage{graphicx}  
\usepackage{dcolumn}   
\usepackage{bm}        
\usepackage{amssymb}   
\usepackage{amsmath}   
\usepackage{bbm}
\usepackage{draftcopy}
\usepackage{relsize} 
\usepackage{xfrac}    
\usepackage{slashed}

\usepackage{color}
\definecolor{dgreen}{cmyk}{1.,0.,1.,0.2}        
\definecolor{orange}{cmyk}{0.,0.353,1.,0.}    

\usepackage[bookmarks]{hyperref}

\newcommand{\di}{{\rm d}}

\newcommand{\be}{\begin{equation}}
\newcommand{\ee}{\end{equation}}                                                                               
\newcommand{\bea}{\begin{eqnarray}}
\newcommand{\eea}{\end{eqnarray}}

\begin{document}
\title{Self-consistent thermodynamic potential for magnetized QCD matter}

\author{Gaoqing Cao}
\affiliation{School of Physics and Astronomy, Sun Yat-sen University, Zhuhai 519088, China}
\author{Jianing Li}
\affiliation{Physics Department, Tsinghua University, Beijing 100084, China}

\date{\today}

\begin{abstract}
Within the two-flavor Nambu--Jona-Lasinio model, we derive a self-consistent thermodynamic potential $\Omega$ for a QCD matter in an external magnetic field $B$. To be consistent with Schwinger's renormalization spirit, counter terms with vacuum quark mass are introduced into $\Omega$ and then the explicit $B$-dependent parts can be regularized in a cutoff-free way. Following that, explicit expressions of gap equation and magnetization can be consistently obtained according to the standard thermodynamic relations. The formalism is able to reproduce the paramagnetic feature of a QCD matter without ambiguity. For more realistic study, a running coupling constant is also adopted to account for the inverse magnetic catalysis effect. It turns out that the running coupling would greatly suppress magnetization at large $B$ and is important to reproduce the temperature enhancement effect to magnetization. The case with finite baryon chemical potential is also explored: no sign of first-order transition is found by varying $B$ for the running coupling and the de Haas-van Alphen oscillation shows up in the small $B$ region. 
\end{abstract}


\maketitle

\section{Introduction}
Extremely strong magnetic fields could be produced in peripheral relativistic heavy ion collisions (HICs)~\cite{Skokov:2009qp,Deng:2012pc} and is also expected to exist in magnetars~\cite{Duncan:1992hi,Thompson:1993hn,Olausen:2013bpa} and the early Universe~\cite{Vachaspati:1991nm,Baym:1995fk,Grasso:2000wj}. For that considerations, a lot of work has been carried out to understand the systematic features of quantum chromodynamics (QCD) matter under an external magnetic field. One important aspect is to study QCD phase transition in a strong magnetic field: as the magnitude of magnetic field is of the order of the QCD energy scale $\Lambda_{\rm QCD}\sim0.2\,{\rm GeV}$, the effect is expected to be considerable. In the end of 20th century, experts took the magnetic field into account in the chiral effective Nambu--Jona-Lasinio model and established the basic notion of "magnetic catalysis effect" to chiral condensate~\cite{Klimenko:1990rh,Klimenko:1992ch,Klimenko:1991he,Klevansky:1992qe,Gusynin:1994re,Gusynin:1994xp}. However,  in 2012, the first-principle lattice QCD (LQCD) simulations~\cite{Bali:2011qj,Bali:2012zg} showed that the chiral condensate could decrease with larger magnetic field at the pseudo-critical temperature $T\sim 0.155\,{\rm GeV}$, known as "inverse magnetic catalysis effect". Such anomalous feature had drawn most attentions of researchers interested in the thermodynamic properties of QCD matter and the QCD phase has been widely explored in the circumstances where magnetic fields are involved, refer to the reviews Ref.~\cite{Miransky:2015ava,Andersen:2014xxa,Cao:2021rwx} and the literatures therein. Specially, it is of great interest that charged pion superfluidity and rho superconductivity were found to be possible in the QCD system under parallel magnetic field and rotation~\cite{Liu:2017spl,Cao:2019ctl,Cao:2020pmm}.

Besides, magnetization is also an important thermodynamic quantity to understand QCD matter. As early as 2000, the magnetization had already been briefly explored as one aspect of magnetic oscillation phenomena in finite density quark matter~\cite{Ebert:1999ht,Vdovichenko:1999hc}. In 2013, both the hadron resonance gas model~\cite{Endrodi:2013cs} and LQCD~\cite{Bali:2013esa,Bonati:2013lca} had been adopted to study the magnetization and the results turned out that the QCD matter is consistently paramagnetic at zero temperature. The $2+1$ LQCD simulations had been extended to finite temperature the next year and the magnetization was found to be enhanced by thermal motions~\cite{Bali:2013owa}. In the following years, only few works concerned the magnetization feature in chiral models such as the two-flavor chiral perturbation theory~\cite{Hofmann:2021bac,Hofmann:2020lfp}, three-flavor Polyakov-linear-sigma (PLS) model~\cite{Tawfik:2017cdx}, and two- and three-flavor (Polyakov-)NJL model~\cite{Avancini:2020xqe,Tavares:2021fik}. The studies in PLS and (P)NJL models seem more realistic as chiral symmetry breaking and restoration were self-consistently taken into account for the evaluation of magnetization. However, compared to previous thermodynamic potential~\cite{Ebert:1999ht}, it is unsatisfied that one had to introduce a cutoff for the explicitly magnetic field dependent terms to evaluate magnetization in the PNJL model~\cite{Avancini:2020xqe}. Furthermore, the definition of magnetization seemed ambiguous as one must additionally apply the renormalization scheme of LQCD simulations~\cite{Bali:2013esa} to get the correct paramagnetic feature~\cite{Avancini:2020xqe}. That is not self-consistent as it seems that the expressions of gap equation and magnetization are not derived from the same thermodynamic potential.

This work is devoted to solving the regularization problem of (P)NJL model in a self-consistent way. In Sec.\ref{formalism}, we will derive a self-consistent thermodynamic potential for finite magnetic field, temperature and baryon chemical potential. From that, expressions of gap equation and magnetization can be given explicitly according to thermodynamic relations. Then, numerical calculations will be carried out in Sec.\ref{numerical}, where we compare the results with different regularization schemes or different forms of coupling constants. Finally, we summarize in Sec.\ref{summary}.

\section{The self-consistent formalism}\label{formalism}
The Lagrangian density of the two-flavor NJL model with baryon chemical potential $\mu_{\rm B}$ can be given as~\cite{Klevansky:1992qe,Hatsuda:1994pi}
\begin{eqnarray}
{\cal L}=\bar\psi\!\left[i\slashed{\cal D}\!-\!i\gamma^4{\mu_{\rm B}\over 3}\!-\!m_0\right]\!\psi\!+\!G(eB)\!\left[\left(\bar\psi\psi\right)^2\!+\!\left(\bar\psi i\gamma_5\boldsymbol\tau\psi\right)^2\right]\label{Lag}
\end{eqnarray}
in Euclidean space, where $\psi=(u,d)^T$ represents the two-flavor quark field, $m_0$ is its current mass, and ${\boldsymbol{\tau}}$ are Pauli matrices in flavor space. In minimal coupling scheme, the covariant derivative is defined as $D_\mu\equiv\partial_\mu-iqA_\mu$ with the electric charge matrix $q\equiv{\rm diag}(q_{\rm u},q_{\rm d})={\rm diag}({2\over3},-{1\over3})e$ and the magnetic effect introduced through the vector potential $A_\mu$. For more general consideration, we have introduced a coupling constant $G(eB)$ that could run with the magnetic field $B$ here.

To obtain the analytic form of the basic thermodynamic potential, we take Hubbard-Stratonovich transformation with the help of the auxiliary fields $\sigma=-2{G}\bar{\psi}\psi$ and ${\boldsymbol{\pi}}=-2{G}\bar{\psi}i\gamma^5{\boldsymbol{\tau}}\psi$~\cite{Klevansky:1992qe} and the Lagrangian becomes
\begin{eqnarray}
{\cal L}=\bar{\psi}\!\left[i{\slashed {\cal D}}\!-\!i\gamma^4{\mu_{\rm B}\over 3}\!-\!i\gamma^5\boldsymbol{\tau}\cdot\boldsymbol{\pi}\!-\!\sigma\!-\!m_0\right]\!\psi-{\sigma^2+\boldsymbol{\pi}^2\over4G(eB)}.
\end{eqnarray}
We assume $\langle\sigma\rangle\equiv m-m_0\neq0$ and $\langle\boldsymbol{\pi}\rangle=0$ in mean field approximation, and then the quark degrees of freedom can be integrated out to give the thermodynamic potential formally as
\bea
\Omega={(m-m_0)^2\over4G(eB)}-{T\over V}{\rm Tr}\ln\left[i{\slashed {\cal D}}-m-i\gamma^4{\mu_{\rm B}\over 3}\right]
\eea
with the trace ${\rm Tr}$ over the coordinate, spinor, flavor and color spaces. Recalling that the quark propagator in a magnetic field takes the form ${\cal S}=-\left[i{\slashed {\cal D}}-m-i\gamma^4{\mu_{\rm B}\over 3}\right]^{-1}$, $\Omega$ can be alternatively presented as 
\bea
\Omega={(m-m_0)^2\over4G(eB)}-{T\over V}\int \di\,m\, {\rm Tr}\,{\cal S}.\label{OmgS}
\eea
Note that the integral limits of $m$ are not important in the second term, because the possible contributions from the lower integral limit is only $B$ dependent which would be definitely fixed by applying Schwinger's renormalization spirit in the following.

At zero temperature and chemical potential, the full fermion propagator in a magnetic field had been well evaluated with the help of proper time by Schwinger in 1951. In coordinate space, it takes the from~\cite{Schwinger:1951nm}:
\begin{widetext}
\begin{eqnarray}
	{\cal S}_{\rm f}(x,x')
	&=&{-i\,q_{\rm f}B\over(4\pi)^2}\int_0^\infty {\di s\over s}\;e^{-iq_{\rm f}\int_{x'}^xA\cdot dx}\exp\Big\{-im^{2}s+{i\over4}\left[{q_{\rm f}B\over\tan(q_{\rm f}Bs)}(y_1^2+y_2^2)+{1\over s}(y_3^2+y_4^2)\right]\Big\}\nonumber\\
	&&\left\{m\!-\!{q_{\rm f}B\over2}\Big[\big(\cot(q_{\rm f}Bs)\gamma^1\!+\!\gamma^2\big)y_1\!+\!\big(\cot(q_{\rm f}Bs)\gamma^2-\gamma^1\big)y_2\Big]\!-\!{1\over2s}\Big[\gamma^3y_3+\gamma^4y_4\Big]\right\}\Big[\cot(q_{\rm f}Bs)
	+\gamma^1\gamma^2\Big]\label{prop_x}
	\end{eqnarray}
with $y_\mu=x_\mu-x_\mu'$ and $s$ the proper time. For the calculation of $\Omega$, the Schwinger phase term $e^{-iq_{\rm f}\int_{x'}^xA\cdot dx}$ is irrelevant since we would take the limit $x\rightarrow x'$. After dropping this term, the left effective propagator becomes translation invariant and can be conveniently presented in energy-momentum space as
\begin{eqnarray}
\hat{{\cal S}}_{\rm f}({p})
&=&i\int_0^\infty {\di s}\exp\Big\{-i (m^{2}+{p}_4^2+p_3^2)s-i{\tan(q_{\rm f}Bs)\over q_{\rm f}B}(p_1^2+p_2^2)\Big\}\left[m-\gamma^4p_4\!-\!\gamma^3p_3\!-\!\gamma^2(p_2+{\tan(q_{\rm f}Bs)}p_1)\right.\nonumber\\
&&\left.-\gamma^1(p_1-{\tan(q_{\rm f}Bs)}p_2)\right]\Big[1
+{\gamma^1\gamma^2\tan(q_{\rm f}Bs)}\Big].\label{prop_p}
\end{eqnarray}
In vanishing $B$ limit, the well-known fermion propagator ${\cal S}({p})={1\over m-\slashed{p}}$ can be reproduced by completing the integration over $s$, hence the effective propagator is helpful for the discussion of regularization. Then, the bare thermodynamic potential follows directly as
\bea
\Omega_0=\!\!{(m-m_0)^2\over4G(eB)}\!+\!{N_{\rm c}\over8\pi^2}\!\sum_{\rm f=u,d}\int_0^\infty {\di s\over s^3}\,e^{-m^{2}s}{q_{\rm f}Bs\over\tanh(q_{\rm f}Bs)}\label{Omg0}
\eea
after substituting the propagator Eq.\eqref{prop_x} into Eq.\eqref{OmgS}. 
 
 The last term of Eq.\eqref{Omg0} is divergent and must be regularized for exploring physics. If we formally expand it as a serial sum of $B^{2k}~({k\in\mathbb{N}})$ around $B\sim0$, we would find that only the $B^0$ and $B^2$ terms are divergent. According to Schwinger's initial proposal~\cite{Schwinger:1951nm}, the $B^0$ term is physics irrelevant and the $B^2$ terms can be absorbed by performing renormalizations of electric charges and magnetic field. Then, the finite form of Eq.\eqref{Omg0} would be
\bea
\Omega_0={(m-m_0)^2\over4G(eB)}+{N_{\rm c}\over8\pi^2}\sum_{\rm f=u,d}\int_0^\infty {\di s\over s^3}\,e^{-m^{2}s}\left[{q_{\rm f}Bs\over\tanh(q_{\rm f}Bs)}-1-{1\over3}(q_{\rm f}Bs)^2\right].
\eea
This is correct when the magnetic field is much smaller than the current mass square $m^2$ in QED systems. But for QCD systems, the dynamical mass $m$ is itself determined by the minimum of the thermodynamic potential, the $B^0$ term can not be dropped at all~\cite{Ebert:1999ht}. Moreover, the dynamical mass $m$ is also $B$-dependent due to magnetic catalysis effect~\cite{Gusynin:1994xp}, the term $e^{-m^{2}s}{1\over3}(q_{\rm f}Bs)^2$ actually contains $o(B^4)$ terms which can not be absorbed by the renormalizations of electric charges and magnetic field. 

The solutions could be the following. Firstly, the $B^0$ term can be recovered with three momentum cutoff according to the discussions in Ref.~\cite{Ebert:1999ht}, then we have
\bea
\Omega_0={(m-m_0)^2\over4G(eB)}+{N_{\rm c}\over8\pi^2}\sum_{\rm f=u,d}\int_0^\infty {\di s\over s^3}\,e^{-m^{2}s}\left[{q_{\rm f}Bs\over\tanh(q_{\rm f}Bs)}-1\right]-4N_c\int^\Lambda{\di ^3p\over(2\pi)^3} E_{\rm p}(m)
\eea
with $E_{\rm p}(m)=(p^2+m^2)^{1/2}$. Next, to absorb the $B^2$ divergent term but not $o(B^4)$ terms, we could refer to the term with vacuum quark mass $m_{\rm v}$ for help. Then, a thermodynamic potential consistent with Schwinger's renormalization spirit can be given as
\bea
\Omega_0&=&{(m-m_0)^2\over4G(eB)}-4N_c\int^\Lambda{\di ^3p\over(2\pi)^3} E_{\rm p}(m)+{N_{\rm c}\over8\pi^2}\sum_{\rm f=u,d}\int_0^\infty {\di s\over s^3}\,\left(e^{-m^{2}s}-e^{-m_{\rm v}^{2}s}\right)\left[{q_{\rm f}Bs\over\tanh(q_{\rm f}Bs)}-1\right]\nonumber\\
&&+{N_{\rm c}\over8\pi^2}\sum_{\rm f=u,d}\int_0^\infty {\di s\over s^3}\,e^{-m_{\rm v}^{2}s}\left[{q_{\rm f}Bs\over\tanh(q_{\rm f}Bs)}-1-{1\over3}(q_{\rm f}Bs)^2\right].
\eea
Note that the subtracted term with integrand $e^{-m_{\rm v}^{2}s}{1\over3}(q_{\rm f}Bs)^2$ only contains $B^2$ term as $m_{\rm v}$ is a constant. 

Eventually, to make sure the pressure to be consistent with the one given in Ref.~\cite{Schwinger:1951nm} when $m=m_{\rm v}$ for any $B$, $m$-independent terms can be subtracted to get the physical thermodynamic potential as
\bea
\Omega_0&=&{(m-m_0)^2-(m_{\rm v}-m_0)^2\over4G(eB)}-4N_c\int^\Lambda{\di ^3p\over(2\pi)^3} [E_{\rm p}(m)-E_{\rm p}(m_{\rm v})]+{N_{\rm c}\over8\pi^2}\sum_{\rm f=u,d}\int_0^\infty {\di s\over s^3}\,\left(e^{-m^{2}s}-e^{-m_{\rm v}^{2}s}\right)\nonumber\\
&&\times\left[{q_{\rm f}Bs\over\tanh(q_{\rm f}Bs)}-1\right]+{N_{\rm c}\over8\pi^2}\sum_{\rm f=u,d}\int_0^\infty {\di s\over s^3}\,e^{-m_{\rm v}^{2}s}\left[{q_{\rm f}Bs\over\tanh(q_{\rm f}Bs)}-1-{1\over3}(q_{\rm f}Bs)^2\right].
\eea
This form of $\Omega_0$ would be adopted for analytic derivations in the following and numerical calculations in next section. Finite temperature and chemical potential usually do not induce extra divergence and the corresponding terms of thermodynamic potential can be easily evaluated with the help of Landau levels as
\bea
\Omega_{T\mu}&=&-2N_{\rm c}T\sum_{\rm f=u,d}^{\rm t=\pm}{|q_{\rm f}B|\over 2\pi}\sum_{n=0}^\infty\alpha_{\rm n}\int_{-\infty}^\infty {\di  p_3\over2\pi}\ln\left[1+e^{-{1\over T}\left(E_{\rm f}^{\rm n}(p_3,m)+t{\mu_{\rm B}\over 3}\right)}\right],
\eea
where $\alpha_{\rm n}=1-\delta_{\rm n0}/2$ and $E_{\rm f}^{\rm n}(p_3,m)=(2n|q_{\rm f}B|+p_3^2+m^2)^{1/2}$. 
So the total thermodynamic potential of a magnetized QCD matter is $\Omega=\Omega_0+\Omega_{T\mu}$, and the expressions of gap equation and magnetization follow the thermodynamic relations ${\partial\Omega/ \partial m}=0$ and ${\cal M}=-{\partial\Omega/ \partial eB}$ as
\bea
0&=&{m-m_0\over2G(eB)}-4N_c\int^\Lambda{\di ^3p\over(2\pi)^3} {m\over E_{\rm p}(m)}-{N_{\rm c}m\over4\pi^2}\sum_{\rm f=u,d}\int_0^\infty {\di s\over s^2}\,e^{-m^{2}s}\left[{q_{\rm f}Bs\over\tanh(q_{\rm f}Bs)}-1\right]+2N_{\rm c}\sum_{\rm f=u,d}^{\rm t=\pm}{|q_{\rm f}B|\over 2\pi}\sum_{n=0}^\infty\alpha_{\rm n}\int_{-\infty}^\infty {\di  p_3\over2\pi}\nonumber\\
&&{m\over E_{\rm f}^{\rm n}(p_3,m)}{1\over 1+e^{{1\over T}\left[E_{\rm f}^{\rm n}(p_3,m)+t{\mu_{\rm B}\over 3}\right]}},\\
{\cal M}&=&{(m-m_0)^2-(m_{\rm v}-m_0)^2\over4}{G'(eB)\over G^2(eB)}-{N_{\rm c}\over8\pi^2}\sum_{\rm f=u,d}\int_0^\infty {\di s\over s^3}\,\left(e^{-m^{2}s}-e^{-m_{\rm v}^{2}s}\right)\left[{\tilde{q}_{\rm f}s\over\tanh(q_{\rm f}Bs)}-{\tilde{q}_{\rm f}q_{\rm f}Bs^2\over\sinh^2(q_{\rm f}Bs)}\right]-\nonumber\\
&&{N_{\rm c}\over8\pi^2}\sum_{\rm f=u,d}\int_0^\infty {\di s\over s^3}\,e^{-m_{\rm v}^{2}s}\left[{\tilde{q}_{\rm f}s\over\tanh(q_{\rm f}Bs)}-{\tilde{q}_{\rm f}q_{\rm f}Bs^2\over\sinh^2(q_{\rm f}Bs)}-{2\over3}\tilde{q}_{\rm f}q_{\rm f}Bs^2\right]+2N_{\rm c}T\sum_{\rm f=u,d}^{\rm t=\pm}{|\tilde{q}_{\rm f}|\over 2\pi}\sum_{n=0}^\infty\alpha_{\rm n}\int_{-\infty}^\infty {\di  p_3\over2\pi}\nonumber\\
&&\ln\left[1+e^{-{1\over T}\left(E_{\rm f}^{\rm n}(p_3,m)+t{\mu_{\rm B}\over 3}\right)}\right]-2N_{\rm c}\sum_{\rm f=u,d}^{\rm t=\pm}{|q_{\rm f}B|\over 2\pi}\sum_{n=0}^\infty\alpha_{\rm n}\int_{-\infty}^\infty{\di  p_3\over2\pi}{n|\tilde{q}_{\rm f}|\over E_{\rm f}^{\rm n}(p_3,m)}{1\over 1+e^{{1\over T}\left[E_{\rm f}^{\rm n}(p_3,m)+t{\mu_{\rm B}\over 3}\right]}}\label{M}
\eea
with $\tilde{q}_{\rm f}={q}_{\rm f}/e$. 

For comparison, the gap equation and magnetization in the so-called {\it vacuum magnetic regularization} (VMR)~\cite{Avancini:2020xqe} are
\bea
0&=&{m-m_0\over2G(0)}-4N_c\int^\Lambda{\di ^3p\over(2\pi)^3} {m\over E_{\rm p}(m)}-{N_{\rm c}m\over4\pi^2}\sum_{\rm f=u,d}\int_0^\infty {\di s\over s^2}\,e^{-m^{2}s}\left[{q_{\rm f}Bs\over\tanh(q_{\rm f}Bs)}-1-{1\over3}(q_{\rm f}Bs)^2\right]\nonumber\\
&&-{N_{\rm c}m\over12\pi^2}\sum_{\rm f=u,d}\int_{1\over\Lambda^2}^\infty {\di s\over s^2}\,e^{-m^{2}s}(q_{\rm f}Bs)^2,\label{gapVMR}\\
{\cal M}_0&=&-{N_{\rm c}\over8\pi^2}\sum_{\rm f=u,d}\!\int_0^\infty {\di s\over s^3}\,e^{-m^{2}s}\!\left[{\tilde{q}_{\rm f}s\over\tanh(q_{\rm f}Bs)}-{\tilde{q}_{\rm f}q_{\rm f}Bs^2\over\sinh^2(q_{\rm f}Bs)}-{2\over3}\tilde{q}_{\rm f}q_{\rm f}Bs^2\right]-{N_{\rm c}\over12\pi^2}\sum_{\rm f=u,d}\int_{1\over\Lambda^2}^\infty {\di s\over s}\left(e^{-m^{2}s}-e^{-m_{\rm v}^{2}s}\right)\tilde{q}_{\rm f}q_{\rm f}B\nonumber\\\label{M0}
\eea
at zero temperature for a constant coupling $G(0)$. But instead of proper-time regularization~\cite{Avancini:2020xqe}, we regularize the explicitly $B$-independent term with three momentum cutoff for better comparison here. Note that the $m_{\rm v}$-dependent term in Eq.\eqref{M0} is important to reproduce the paramagnetic feature of QCD matter though they did not manage to give the explicit form~\cite{Avancini:2020xqe}. 
\end{widetext}

\section{Numerical results}\label{numerical}
To carry out numerical calculations, the model parameters are fixed as $m_{\rm 0}=5\,{\rm MeV},\, \Lambda=653\,{\rm MeV},\, G(0)\Lambda^2=2.10$ by fitting to the vacuum values: chiral condensate $\langle\bar\psi\psi\rangle=-2\times(250\,{\rm MeV})^3$, pion mass $m_\pi=135\,{\rm MeV}$, and pion decay constant $f_\pi=93\,{\rm MeV}$~\cite{Zhuang:1994dw,Rehberg:1995kh}. Accordingly, the vacuum quark mass is $m_{\rm v}=-2G(0)\langle\bar\psi\psi\rangle+m_0=0.313\,{\rm GeV}$. And the explicit form of $G(eB)$ should be given to study the effect of finite magnetic field. In Ref.~\cite{Cao:2021rwx}, a form of $G(eB)$ had been determined by fitting to the data of $\pi^0$ mass from LQCD simulations, with which we were able to explain inverse magnetic catalysis effect at larger $B$. However, there was nonphysical increasing of G(eB) around $eB\sim0$; to avoid that, we choose to fit to the region $eB \geq 0.6\, {\rm GeV}^2$ here and get a monotonic form $G(eB)={G(0)\over 1+0.524\,eB^2}$. Hence, ${G'(eB)\over G^2(eB)}=-{1.048\,eB\over G(0)}$.

\begin{figure}[!htb]
	\begin{center}
	\includegraphics[width=8cm]{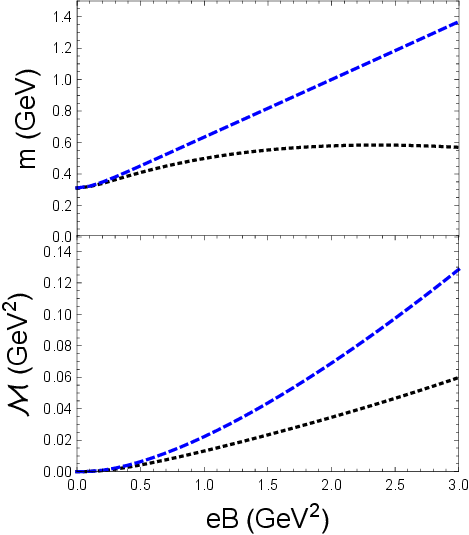}
		\caption{The dynamical mass $m$ (upper panel) and magnetization ${\cal M}$ (lower panel) with the self-consistent regularization (blue dashed lines) and vacuum magnetic regularization (black dotted lines) schemes at zero temperature.}\label{T0}
	\end{center}
\end{figure}

For a constant coupling $G(0)$, we compare the results of our self-consistent regularization scheme with those of VMR scheme in Fig.~\ref{T0} at zero temperature. Both results are consistent with the LQCD data~\cite{Bali:2013esa,Ding:2022tqn} for the region $0\leq eB \leq 0.6\, {\rm GeV}^2$, but they diverge quite much from each other for larger $B$. In our opinion, the cutoff to the explicitly $B$-dependent term in VMR, see the last term in Eq.\eqref{gapVMR}, would introduce artifact at larger $B$ -- the nonmonotonic feature of $m$ is a reflection of that. Here and in the following, one might suspect the legality of studying the effect of magnet field as large as $3\,{\rm GeV}^2$ in the low-energy effective theory such as the NJL model. We would like to remind the readers that the coupling between quarks and magnetic field represents the quantum electrodynamics (QED) interaction, and the effective range of $B$ is infinite according to the renormalizability of the first term on the right-hand side of Eq.\eqref{Lag}~\cite{Schwinger:1951nm}. By adding the four-fermion interaction terms, see the second term on the right-hand side of Eq.\eqref{Lag}, one actually tries to approximate the low-energy QCD with the effective NJL model. If one neglects the interplay between QED and QCD interactions, the effective range of  $B$ would sustain to infinity. However, in reality, there is interplay between the QED and QCD interactions, such as asymptotic freedom with increasing $B$. In our opinion, finding out the interplay composes an important mission of the model study. Of course, to keep the NJL model qualitatively valid for large $B$, one should make sure that the renormalizability of the $B$-dependent part is not affected by the cutoff from the QCD part. That is why it is important to present the self-consistent regularization here. Moreover, the absolute value of chiral condensate was found to linearly increase with large $eB$ at zero temperature in NJL model~\cite{Wang:2017pje}, which is consistent with the results of LQCD up to $eB\sim 2.5~{\rm GeV}^2 (>>\Lambda^2)$~\cite{Ding:2022tqn}. This strongly indicates that the valid range of $B$ could be very large in NJL model once the $B$-dependent part is properly renormalized.

\begin{figure}[!htb]
	\begin{center}
	\includegraphics[width=8cm]{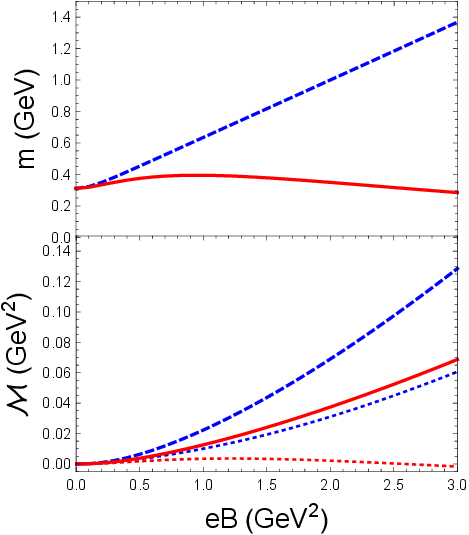}
		\caption{The dynamical mass $m$ (upper panel) and magnetization ${\cal M}$ (lower panel) with the constant coupling $G(0)$ (blue dashed lines) and the running coupling $G(eB)$ (red lines) at zero temperature. The dotted lines correspond to the corresponding contributions of the second term in Eq.\eqref{M}.}\label{T0B}
	\end{center}
\end{figure}

In the following, we would explore how a running coupling constant could affect the dynamical mass and the corresponding magnetization in the self-consistent regularization.  At zero temperature, the results with $G(0)$ and $G(eB)$ are shown together in Fig.~\ref{T0B}. Due to the running of coupling constant, $m$ shows a nonmonotonic feature though the absolute value of chiral condensate, $(m-m_0)/2G(eB)$, increases almost linearly with $B$~\cite{Cao:2021rwx}. Accordingly, the second term in Eq.\eqref{M} demonstrates a nonmonotonic feature and becomes negative at larger $B$. Such feature is responsible for the strong suppression of magnetization at larger $B$ in the case with $G(eB)$ compared to that with $G(0)$.

\begin{figure}[!htb]
	\begin{center}
	\includegraphics[width=8cm]{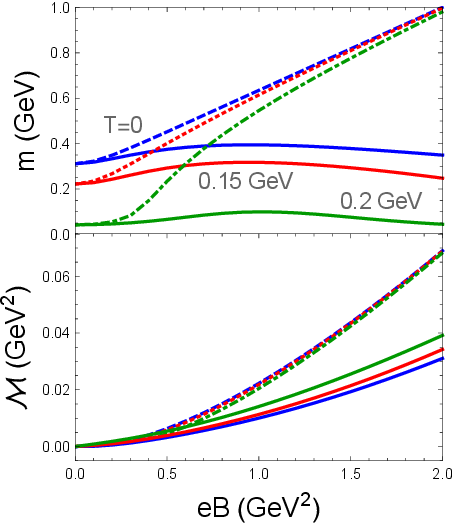}
		\caption{The dynamical mass $m$ (upper panel) and magnetization ${\cal M}$ (lower panel) as functions of the magnetic field $B$ at temperature $T=0, 0.15,$ and $0.2\,{\rm GeV}$. The dashed, dotted, and dashdotted lines correspond to the results with the constant coupling $G(0)$ and the solid lines correspond to the results with the running coupling $G(eB)$.}\label{TB}
	\end{center}
\end{figure}
At finite temperature, the results are illustrated in Fig.~\ref{TB}. As we can see, the temperature tends to suppress magnetization in the case with $G(0)$ but enhance magnetization in the case with $G(eB)$. In their book, Landau and Lifshitz had calculated magnetic susceptibility $\chi\equiv{e{\cal M}\over{\cal N} B}$ of a non-relativistic dilute electronic gas at high temperature and found it decreases as $1/T$~\cite{Landau1999}. To be concrete, the situations they considered are $\sqrt{B}\ll T\ll m_{\rm e}$ and the electric chemical potential $-\mu_{\rm e} (\gtrsim m_{\rm e})$ changes with $T$ to keep the total number ${\cal N}$ constant. If we keep $-\mu_{\rm e} (\gtrsim m_{\rm e})$ a constant, then the total electronic number ${\cal N}$ could be easily evaluated to increase with temperature as $T^{3/2}$. Therefore, the magnetization ${\cal M}=\chi{\cal N} B/e$ would increase with temperature as $\sqrt{T}$, and the result with $G(eB)$ is qualitatively consistent with the non-relativistic study. That is not the end of story: when we keep $m=m_{\rm v}$ for $G(0)$, ${\cal M}$ would increase with $T$ for a given $B$; so it is adequate chiral symmetry restoration induced by $T$ that reduces the contribution of second term in Eq.\eqref{M} and thus reverses the trend. One can refer to Fig.\ref{T0B} for the dynamical mass effect on magnetization. For $G(eB)$, $m$ changes mildly with $B$ for a given $T$, that is, the large mass gaps induced by $T$ at vanishing $B$ sustain to strong magnetic field. According to our analysis, it is the great enhancement of the forth $T$-dependent term in Eq.\eqref{M} that helps to recover the trend of naive expectation. In fact, the result with $G(eB)$ is qualitatively consistent with that found in LQCD simulations at finite temperature~\cite{Bali:2013owa}, so we conclude that the running coupling is able to consistently explain both inverse magnetic catalysis effect and magnetization enhancement with temperature. 
\begin{figure}[!htb]
	\begin{center}
	\includegraphics[width=8cm]{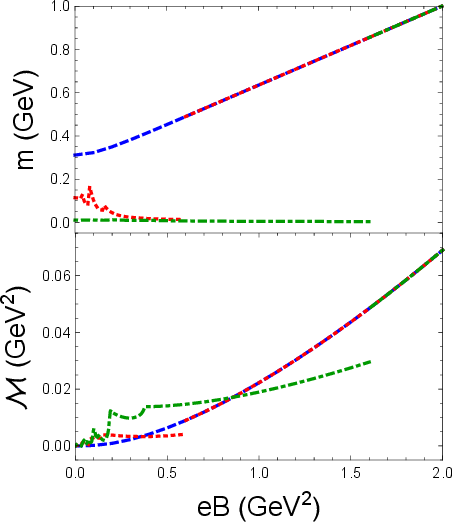}
		\caption{The dynamical mass $m$ (upper panel) and magnetization ${\cal M}$ (lower panel) as functions of the magnetic field $B$ at baryon chemical potential $\mu_{\rm B}=0, 1,$ and $1.5\,{\rm GeV}$. The dashed, dotted, and dashdotted lines correspond to the results with the constant coupling $G(0)$.}\label{muB0}
	\end{center}
\end{figure}
\begin{figure}[!htb]
	\begin{center}
	\includegraphics[width=8cm]{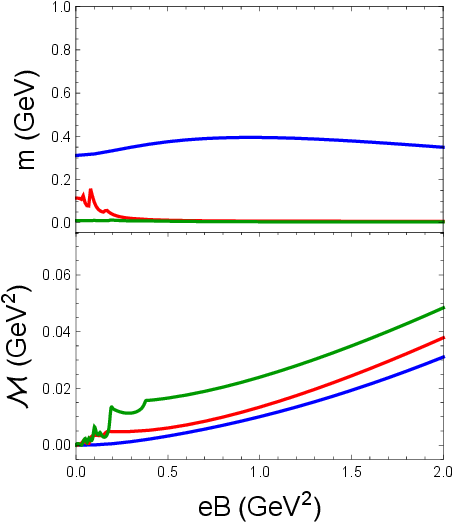}
		\caption{The dynamical mass $m$ (upper panel) and magnetization ${\cal M}$ (lower panel) as functions of the magnetic field $B$ at baryon chemical potential $\mu_{\rm B}=0, 1,$ and $1.5\,{\rm GeV}$. The solid lines correspond to the results with the running coupling $G(eB)$.}\label{muB}
	\end{center}
\end{figure}

At finite baryon chemical potential, the results are illustrated in Fig.~\ref{muB0} and Fig.~\ref{muB}. For $G(0)$, $m$ always changes discontinuously with $B$ for $\mu_{\rm B}>m_{\rm v}$, which signals a first-order transition. But for $G(eB)$, $m$ only changes slightly at $\mu_{\rm B}=0$ and no sign of first-order transition could be identified for a given $\mu_{\rm B}$. The de Haas-van Alphen oscillation~\cite{Landau1999} can be found both in the evolutions of $m$ and ${\cal M}$ with $B$: the effect is significant to $m$ only when $\mu_{\rm B}$ is a little larger than $3m_{\rm v}$ but is significant to ${\cal M}$ for any $\mu_{\rm B}>3m_{\rm v}$. According to the mechanism of de Haas-van Alphen oscillation~\cite{Landau1999}, the last non-analytic points of ${\cal M}$ can be roughly determined by $\sqrt{2q_{\rm d}|B|}\approx\mu_{\rm B}/3$, that is, $eB\approx 0.167\,{\rm GeV}^2$ for $\mu_{\rm B}=1\,{\rm GeV}$ and $eB\approx 0.375\,{\rm GeV}^2$ for $\mu_{\rm B}=1.5\,{\rm GeV}$. That is consistent with the numerical results shown in the lower panel of Fig.~\ref{muB}. Moreover, at larger $B$, ${\cal M}$ does not depend on $\mu_{\rm B}$ for $G(0)$ due to the "Silver braze" property but increases with $\mu_{\rm B}$ for $G(eB)$ due to the strong suppression of $m$.

\section{Summary}\label{summary}
In this work, a self-consistent thermodynamic potential has been obtained for a magnetized QCD matter in two-flavor NJL model by following Schwinger's renormalization spirit. The thermodynamic potential is free of cutoff for the explicitly magnetic field dependent terms, and explicit expressions of gap equation and magnetization could be derived from that by following the thermodynamic relations. Compared to the VMR scheme, the numerical calculations showed that magnetic catalysis effect persists to very large magnetic field at zero temperature when adopting the self-consistent scheme, and the magnetization is strongly affected accordingly.

Within the self-consistent scheme, results with the constant coupling $G(0)$ and running coupling $G(eB)$ are compared with each other. At zero temperature and chemical potential, the running coupling greatly suppresses the dynamical mass $m$ at large magnetic field $B$ and thus reduces the magnetization ${\cal M}$ a lot. At finite temperature $T$, ${\cal M}$ decreases with $T$ for $G(0)$ due to adequate suppression of $m$ but increases with $T$ for $G(eB)$ due to the persistence of large mass gaps at large $B$. At finite baryon chemical potential $\mu_{\rm B}$, no sign of first-order transition could be identified for $G(eB)$ by varying $B$ and the de Haas-van Alphen oscillation could be found both in the evolutions of $m$ and ${\cal M}$ with $B$.

Since we found that the regularization scheme could affect the result greatly in the large magnetic field region,  we would try to perform similar study in three-flavor NJL or PNJL model in the future. Then, we could compare the evaluations of magnetization with the LQCD data in the region $0\leq eB\leq 1\,{\rm GeV^2}$ for finite temperature~\cite{Bali:2013owa} and give further predictions for much larger magnetic field. The situation with finite baryon chemical potential could also be explored for completeness, which might help us to understand the properties of magnetars.

\section*{Acknowledgments}
G.C. is supported by the National Natural Science Foundation of China with Grant No. 11805290. J. Li is supported by the National Natural Science Foundation of China with Grant No. 11890712.

\end{document}